\documentclass[journal]{IEEEtran}
\IEEEoverridecommandlockouts
\usepackage{bbding}
\usepackage{url}
\usepackage{booktabs}

\usepackage{CJK}
\usepackage{subfigure}
\usepackage{amsmath,amssymb,amsthm}
\usepackage{multirow}
\usepackage{fancybox}
\usepackage[dvipsnames,svgnames,x11names]{xcolor}
\usepackage{graphicx}
\usepackage{bbm}
\usepackage{algorithm} 
\usepackage{algorithmic} 
\usepackage{verbatim}
\usepackage{booktabs}
\usepackage[UKenglish]{babel}
\usepackage{setspace}
\usepackage{color}
\makeatletter
\newcommand*{\rom}[1]{\expandafter\@slowromancap\romannumeral #1@}
\makeatother
\newcommand\norm[1]{\left\lVert#1\right\rVert}

\DeclareMathOperator*{\argmax}{argmax}

\ifCLASSINFOpdf
\else
\fi
\begin{document}
%
\title{Placement of EV Charging Stations---Balancing Benefits among Multiple Entities}
%
%
%

\author{Chao~Luo,
        Yih-Fang~Huang,~\IEEEmembership{Fellow,~IEEE,}
        and~Vijay~Gupta,~\IEEEmembership{Member,~IEEE}
\thanks{C. Luo, Y.-F. Huang, and V. Gupta are with the Department
of Electrical Engineering, University of Notre Dame, Notre Dame,
IN, 46556 USA e-mail: \{cluo1, huang, vgupta2\}@nd.edu }.
\thanks{This paper was presented, in part, at the 2015 IEEE 81st Vehicular Technology Conference. This work has been partially supported by the National Science Foundation under grants CNS-1239224 and ECCS-0846631.}
}

\maketitle

\begin{abstract}
This paper studies the problem of multi-stage placement of electric vehicle (EV) charging stations with incremental EV penetration rates. A nested logit model is employed to analyze the charging preference of the individual consumer (EV owner), and predict the aggregated charging demand at the charging stations. The EV charging industry is modeled as an oligopoly where the entire market is dominated by a few charging service providers (oligopolists). At the beginning of each planning stage, an optimal placement policy for each service provider is obtained through analyzing strategic interactions in a Bayesian game. To derive the optimal placement policy, we consider both the transportation network graph and the electric power network graph. A simulation software---The EV Virtual City 1.0---is developed using Java to investigate the interactions among the consumers (EV owner), the transportation network graph, the electric power network graph, and the charging stations. Through a series of experiments using the geographic and demographic data from the city of San Pedro District of Los Angeles, we show that the charging station placement is highly consistent with the heatmap of the traffic flow. In addition, we observe a spatial economic phenomenon that service providers prefer clustering instead of separation in the EV charging market.
\end{abstract}


%
\IEEEpeerreviewmaketitle

\section{Introduction}
%
%
%
%
The continued technological innovations in battery and electric drivetrain have made electric vehicles (EVs) a viable solution for a sustainable transportation system. Currently, most EV charging is done either at residences, or for free at some public charging infrastructure provided by municipalities, office buildings, etc. As the EV industry continues to grow, commercial charging stations will need to be strategically added and placed. Development of effective management and regulation of EV charging infrastructure needs to consider the benefits of multiple constituencies---consumers, charging station owners, power grid operators, local government, etc. In this paper, we concentrate on striking a balance among the profits of charging station owners, consumer satisfaction, and power grid's reliability.

Our work is motivated by the desire of service providers to make a forward-looking decision on charging station placement to obtain a good return on their investment. We use the most up-to-date information (i.e., travel pattern, traffic flow, road network, power grid, etc.) to make the best-effort decisions on charging station placement, hoping that service providers will have a good chance to profit over the next few years. In this paper, we do not consider factors such as uncertainties in fuel prices, climate change, population migration etc., which are random and unpredictable. Instead, we assume that some revenue management techniques (i.e., real-time pricing) may be applied to deal with the potential effects of these factors.

We assume that the service providers aim to strike a balance between the competing goals of maximizing their profits and minimizing the disturbance to the electric power network due to large-scale EV charging. Accordingly, we construct a utility function that incorporates both of these aims. Each charging service provider attempts to maximize his/her own expected utility function while satisfying the Quality-of-Service (QoS) constraints through choosing the optimal locations of charging stations that s/he owns. The nested logit model is used to analyze and predict the charging preference of EV owners. At the beginning of each stage, the service providers predict the charging demand of each charging station candidate using the nested logit model. The optimal placement strategy is obtained through a Bayesian game. As the EV penetration rate increases, the existing charging stations may no longer satisfy the QoS constraints and a new stage shall be initiated to place more charging stations.

There is a growing literature addressing the issues relevant to EV charging station placement. \cite{Shaoyun}-\cite{Zonggen} formulated charging station placement as an optimization problem. However, they did not take into account the overall consumer satisfaction and the impact of  EV charging on the electric power network in their works. Besides, their optimization models were formulated from the perspective of a central urban planner rather than that of service providers in a free competitive market.  In \cite{cite8}, the authors presented a strategy to deploy charging stations by analyzing the patterns of residential EV ownership and driving activities. In their work,  they deploy the new charging stations either randomly with no weight or only based on the weights of population. They did not consider the mobility of EVs and the overall consumer experience. In \cite{cite9}, Bernardo \emph{et al.} proposed a discrete choice model (DCM) based framework to study the optimal locations for fast charging stations. They treat each charging station as a player in a noncooperative game. However, the underlying assumption in their work is that each player has complete information about other players, which may be overly restrictive and infeasible in a practical competitive market. In this paper, we propose a Bayesian game framework that does not require the complete information of other players.

The main contributions of our work are: (1) A multi-stage charging station placement strategy with incremental EV penetration rates is first formulated, which takes into account the interactions among EVs,  road network,  and the electric power grid; (2) The nested logit model is then employed to characterize the overall consumer satisfaction and predict the aggregated charging demand, which provides insights into the preferences and decision-making processes of EV owners; (3) An oligopolistic market model of EV charging service providers is studied and a Bayesian game framework is applied to analyze the strategic interactions among service providers. (4) A simulation software has been developed to analyze the interplay among EV owners, road network, power grid, urban infrastructure and charging stations \footnote{The EV Virtual City 1.0 simulator can be downloaded from https://github.com/chaoluond/EVVirtualCity}.

The paper is organized as follows: Section \rom{2} presents the problem formulation. Section \rom{3} discusses the nested logit model and how charging demand is calculated. In Section \rom{4}, we describe the impact of EV charging on the power grid. In Section \rom{5}, a Bayesian game is used to characterize the competition among service providers, and the optimal station placement policy is obtained. Section \rom{6} shows the architecture and applications of the simulation software and discusses a case study in San Pedro District. Conclusions are given in Section \rom{7}. Table \rom{1} provides a full description of parameters and symbols used in the paper.

\section{Problem Formulation}

\begin{table}[htbp]
\center
\caption{Parameter Description}
 \begin{tabular}{lll}
  \toprule
  Parameter & Description & Unit\\
  \midrule
$L$ & Total Candidate Locations & - \\
$N$ & Total EVs & -\\
$\psi_{j,k}^n$ & Charging Demand & kWh\\
$s_{j,k}$ & Placement Indicator & -\\
$F_{j,k}$ & Setup Cost & \$\\
$R_k$ & Total Revenue & \$\\
$c_{j,k}$ & Locational Marginal Price & \$\\
$\Pi_k$ & Total Profit & \$\\
$U_k$ & Overall utility & \$\\
w &  Coef. of EV Charging Penalty & -\\
$\Upsilon_k$ & Average Service Probability & -\\
$\Xi_k$ & Average Service Coverage & -\\
$t_k$ & Average Charging Time & min\\
$p_k$ & Retail Charging Price & \$/kWh\\
$i_n$ & Income of the $n$th EV owner & \$\\
$d_{j,k}^n$ & Deviating distance & km\\
$z_{j,k}^n$ & Indicator of Destination & -\\
$d_{th}$ & Distance threshold & km\\
$r_{j,k}$ & Indicator of Restaurant & - \\
$g_{j,k}$ & Indicator of Shopcenter & -\\
$m_{j,k}$ & Indicator of Supermarket & -\\
$\mathbf{P}_{\textrm{g}}^{\textrm{base}}$ & Active power vector without EV & -\\
$\mathbf{P}_{\textrm{g}}^{\textrm{EV}}$ & Active power vector with EV & -\\
$\mathbf{Q}_{\textrm{g}}^{\textrm{base}}$ & Reactive power vector without EV & -\\
$\mathbf{Q}_{\textrm{g}}^{\textrm{EV}}$ & Reactive power vector with EV & -\\
$v_i$ & Voltage at bus $i$ & Volts\\
$\phi_{ik}$ & Voltage angle between bus $i$, $k$ & Radian\\
$\alpha$ & Coef. of $t_k$ & -\\
$\beta$ & Coef. of $p_k/i_n$ & -\\
$\mu_k$ & Coef. of $d_{j,k}$ & -\\
$\eta_k$ & Coef. of $z_{j,k}$ & -\\
$\gamma_k$ & Coef. of $r_{j,k}$ & -\\
$\lambda_k$ & Coef. of $g_{j,k}$ & -\\
$\delta_k$ & Coef. of $m_{j,k}$ & -\\
\bottomrule
 \end{tabular}
\end{table}

In this paper, we postulate the problem of EV charging with an oligopolistic market structure that has multiple charging service providers (oligopolists). The service providers aim to maximize their expected utility while satisfying the QoS constraints by selecting optimal station placements.

In particular, we consider the case of three service providers that offer three EV charging services \cite{specification}, namely, Level 1, Level 2, and Level 3 (see Table \ref{ta1} for details). Level 1 and Level 2 are AC charging. Level 3 charging is DC Fast charging. Let $\mathcal{O}=\{1,2,3\}$ denote the set of charging service providers. Moreover, we assume that service provider 1 offers Level 1 charging, service provider 2 offers Level 2 charging, and service provider 3 offers Level 3 charging. The three charging levels represent three charging services, which have different charging voltages and currents, charging times, and charging experiences. In economic terms, they are imperfect substitutes to each other. In our model, we are interested in investigating how the different charging services compete with each other in choosing locations and prices. Each service provider can run multiple charging stations. At each planning stage, service providers select some charging stations from a given set of candidates, denoted as $\mathcal{I}=\{1,2,3\cdots,L\}$. The set of EVs is denoted as $\mathcal{E}=\{1,2,3,\cdots,N\}$.

\subsection{The Profit of EV Charging}

\begin{table*}[htbp]
\center
\caption{Charge Method Electrical Ratings}\label{ta1}
 \begin{tabular}{cccc}
  \toprule
  Charging Method& Nominal Supply Voltage(Volts)&Max Current (Amps) & Time from fully depleted to
fully charged \\
  \midrule
 Level 1 &120 vac, 1-phase &12 A & 16-18 hours \\
 Level 2 & 208 to 240 vac, 1-phase&32 A & 3-8 hours \\
 Level 3&600 vdc maximum&400 A maximum & less than 30 minutes  \\
  \bottomrule
 \end{tabular}
\end{table*}

 We assume that service providers run the charging stations like ``chain stores", so that charging stations affiliated with the same service provider have the same retail charging price. The charging stations purchase the electricity from the wholesale market at the locational marginal price (LMP). In a deregulated electricity market (like PJM, NYISO, NEISO, MISO, ERCOT, California ISO in USA,  the New Zealand and Singapore markets), LMP is computed at every node (bus) by the market coordinator. LMP primarily consists of three components: system energy price, transmission congestion cost, and cost of marginal losses \cite{PJM}. Let $p_k$ represent the retail charging price of provider $k\;(k=1,2,3)$, and $p_{-k}$ be the retail charging prices of the other two service providers except $k$. Let $c_{j,k}$ be the LMP of the $j$th charging station candidate of service provider $k$, and $\psi_{j,k}$ be the predicted charging demand at the $j$th charging station candidate of service provider $k$. The vector $S_k=[s_{1,k},s_{2,k},\cdots, s_{L,k}]^{\textrm{T}}$ represents the placement policy of service provider $k$, where $s_{j,k}\in\{0,1\}$ is an indicator with $s_{j,k}=1$ implying that service provider $k$ will place the $j$th charging station. Let $S_{-k}$ represent the placement policies of the other two service providers, $\theta_{j,k}$ be the placement cost of the $j$th charging station. The total profit of service provider $k$ is given by

\begin{equation}
\Pi_k=p_k\Psi_k^{\textrm{T}}S_k-\textrm{diag}[C_k]\Psi_k^{\textrm{T}}S_k
-\Theta_k^{\textrm{T}}S_k,
\end{equation}
and the total revenue of service provider $k$ is,
\begin{equation}\label{revenue}
R_k=p_k\Psi_k^{\textrm{T}}S_k-\textrm{diag}[C_k]\Psi_k^{\textrm{T}}S_k,
\end{equation}
where $\Psi_k=[\psi_{1,k},\psi_{2,k},\psi_{3,k},\cdots,\psi_{L,k}]^{\textrm{T}}$, $C_k=[c_{1,k},c_{2,k},\cdots,c_{L,k}]^{\textrm{T}}$ and $\Theta_k=[\theta_{1,k},\theta_{2,k},\cdots,\theta_{L,k}]^{\textrm{T}}$. The notation $\textrm{diag}[.]$ is an operator to create a diagonal matrix using the underlying vector, and $[.]^{\textrm{T}}$ is the transpose operation. In Equation (\ref{revenue}), the total revenue from the sales is $p_k\Psi_k^{\textrm{T}}S_k$, the cost of purchasing electricity is $\textrm{diag}[C_k]\Psi_k^{\textrm{T}}S_k$, and the placement cost is $\Theta_k^{\textrm{T}}S_k$.

\subsection{The Disturbance on Power Grid Due to EV Charging}
It is conceivable that the simultaneous large-scale EV charging can disrupt the normal operation of the power grid in terms of frequency variation, voltage imbalance, voltage variation, power loss, etc \cite{impact1}-\cite{impact3}. In a conventional power grid, the generators will be incentivized to cooperatively control the output of real power and reactive power to maintain system stability, perform frequency regulation and voltage regulation. The charging service providers typically cannot participate in such market. Instead, we assume that they are fined in proportion to the ``disturbance" they impose on the grid. Thus, the providers must optimally place the charging stations to mitigate the ``disturbance" to the power grid. Accordingly, the overall utility function of service provider can be defined as:

\begin{equation}\label{omega}
U_k=\Pi_k-wB_k,
\end{equation}
where $\Pi_k$ is the total profits from EV charging. $B_k$ characterizes the penalty arising from large-scale EV charging. The variable $w$ is a weighting coefficient, reflecting the tolerance to the penalty. In the following section, we will further discuss how to develop a proper metric to evaluate the penalty $B_k$. We should note that the weighting factor $w$ in Equation (\ref{omega}) offers a mechanism for the charging service providers to strike a balance between their own profit and the ``stress" their charging adds to the power system.  If $w=0$, the impact of EV charging on the grid is not considered at all, and a non-zero value of $w$ implies some impact on the grid---the larger $w$ is, the larger the impact is.  Generally, if the focus is on the charging provider's profit, a small $w$ is used.  In practice, the value of $w$ needs to be determined, e.g., with the help of heuristic and empirical data.

\subsection{Quality-of-Service Constraints}
We use two quality-of-service (QoS) metrics for the service provider: (1) the average service delay probability $\Upsilon_k$, and (2) the average service coverage $\Xi_k,\;(k=1,2,3)$.

\begin{equation}
\Upsilon_k=\frac{1}{N}\sum_{i=1}^N\upsilon_{i,k},
\end{equation}

\begin{equation}
\Xi_k=\frac{1}{N}\sum_{i=1}^N\xi_{i,k},
\end{equation}
where $\upsilon_{i,k}$ is the average service delay probability for the $i$-th EV owner getting the EV charged at service provider $k$. For the $i$th EV owner, $\upsilon_{i,k}$ is defined as the ratio of the number of delayed charging to the total number of charging attempts; $\xi_{i,k}$ is the average number of accessible Level $k$ charging stations along the route from origin to destination. Notice that $\upsilon_{i,k}$ and $\xi_{i,k}$ are random variables that depend on the travel patterns of all EVs, the urban road network and the charging stations. It is, thus, difficult to use a simple formula to compute them. Instead, we employ Mento Carlo method to estimate those two values.

\subsection{Multi-stage Charging Station Planning Scheme}
 At each planning stage, the service providers aim at solving the following fundamental problem to obtain the optimal placement policy subjected to the QoS constraints.

\begin{equation}\label{obj}
\begin{aligned}
&[S_k^T|S_1^{T-1},S_2^{T-1},S_3^{T-1}]=\\
&\argmax_{\substack{{s_{1,k},\cdots,s_{L,k}}\\{s_{j,k}\in\{0,1\}}}}\left\{\mathbb{E}_{S_{-k}}[U_k]|S_1^{T-1},
S_2^{T-1},S_3^{T-1}\right\},
\end{aligned}
\end{equation}
subject to
\begin{equation}
\Upsilon_k\leq\Upsilon^0,
\end{equation}
\begin{equation}
\Xi_k\leq\Xi^0,
\end{equation}
where $\mathbb{E}_{S_{-k}}[.]$ denotes the expectation over $S_{-k}$, and $S_k^T$ is the placement policy at stage $T$. The variables $\Upsilon^0$ and $\Xi^0$ are the predetermined QoS constraints.

To solve this problem, we are confronted with three principal questions: (1) How to predict the aggregated charging demand $\psi_{j,k}$ at each charging station candidate? (2) How to find an appropriate metric to characterize the impacts of EV charging on the power grid? (3) How to derive the optimal placement policy in a tractable way? For the first question, we assume that the service providers estimate the charging demand using a nested logit model. In Section \rom{3}, we will describe the nested logit model and elaborate on how to use this model to analyze consumer behavior and estimate the charging demand.  In Section \rom{4}, we will discuss how the EV charging may impact the power grid, and propose a metric to assess the impacts of EV charging. For the last question, notice that the optimization problem formulated by Equation (\ref{obj}) is intractable since the optimal placement decision for every service provider also depends on the decisions taken by other service providers. To this end, we employ a Bayesian game model to characterize the strategic interaction and price competition among the service providers. We will calculate the optimal placement strategies and prices in Section \rom{5}.

\section{Charging Demand of EV Charging Station}
In this paper, the aggregated charging demand at a charging station candidate is defined as the sum of the product of the probability that EV owners choose that particular charging station and the electricity required to charge the EVs. The charging behaviors of the EV owners may be influenced by many factors that include the charging price, travel cost, amenities at or near the charging station, the travel purpose, EV owner's income, and so on. We use the nested logit model to characterize the attractiveness of a charging station.

\subsection{Nested Logit Model And Probability of Choice}
The nested logit model is widely used in the analysis and prediction of a consumer's choice from a finite set of choice alternatives \cite{cite11}. The main idea of nested logit model is that a consumer is a utility maximizer. The consumer will choose the product which brings him/her the maximum utility.

In our problem, the utility that the $n$-th EV owner can obtain from choosing charging station $j\;(j=1,2,\cdots,L)$ of service provider $k\;(k=1,2,3)$ is denoted as $U_{j,k}^n=\overline{U}_{j,k}^n+\epsilon_{j,k}^n$, where $\overline{U}_{j,k}^n$ is the observable utility and $\epsilon_{j,k}^n$ is the unobservable utility. The vector of unobservable utility $\epsilon^n=[\epsilon_{1,1}^n,\cdots,\epsilon_{L,1}^n,\epsilon_{1,2}^n,\cdots,\epsilon_{L,2}^n, \epsilon_{1,3}^n,\cdots,\epsilon_{L,3}^n]^{\textrm{T}}$ is assumed to have a generalized extreme value (GEV) distribution with cumulative distribution function (CDF) \cite{cite11}

\begin{equation}\label{cdf}
F(\epsilon^n)=\exp\left(-\sum_{k=1}^3\left(\sum_{l=1}^Le^{-\epsilon_{l,k}^n/\sigma_k}\right)^{\sigma_k}\right),
\end{equation}
where $\sigma_k$ is a measure of the degree of independence in the unobservable utility among the charging stations owned by service provider $k$. For the nested logit model, $\epsilon_{j,k}$ is correlated within each charging level, and uncorrelated across different charging levels.

We can decompose the observable utility $\overline{U}_{j,k}^n$ into two components---the utility of choosing service provider $k$ and the utility of choosing a charging station $j$. In addition, we assume home charging is the ``outside good" in this market \cite{outside1}-\cite{outside2}. Thus, $\overline{U}_{j,k}^n$ for EV owner $n$ can be expressed as

\begin{equation}
\overline{U}_{j,k}^n=\overline{W}_{k}^n+\overline{V}_{j,k}^n,
\end{equation}
where $\overline{W}_{k}^n$ is the observable utility of choosing service provider $k$ (choosing nest $k$), and $\overline{V}_{j,k}^n$ is the observable utility of choosing charging station $j$ given that service provider $k$ has been chosen; $\overline{W}_{k}^n$ and $\overline{V}_{j,k}$ are linear weighted combinations of attributes of the charging stations and the EV owner.

Note that the retail charging price and the charging time are two factors differentiating the three charging services. In addition, we assume the income of EV owners will also play a role in choosing charging services. In contrast to our previous work \cite{mywork}, we use a different formula to calculate $\overline{W}_k^n$ here.

\begin{equation}
\overline {W}_k^n =\alpha \frac{1}{t_k}+\beta \frac{p_k}{i_n},
\end{equation}
where $t_k$, $p_k$ and $i_n$ represent, respectively, the average charging time, the retail charging price, and the income of the $n$-th EV owner; and $\alpha, \beta$ are the corresponding weighting coefficients. This model is similar to Ben-Akiva and Lerman's utility model in their study of public transportation mode \cite{transmode}. The value of $\alpha$ is positive because shorter charging time implies a better charging service experience; therefore, leading to higher utility. The value of $\beta$ is negative because a higher retail charging price results in less utility. However, the retail charging price is divided by the income, which reflects that the retail charging price for the EV owners becomes less important as their income increases. As an ``outside good", the utility of home charging is normalized, i.e. $\overline{W}_0^n=0$.

Furthermore, we define $\overline{V}_{j,k}^n$ as follows,
\begin{equation}
\begin{aligned}
\overline{V}_{j,k}^n=&
\mu_kd_{j,k}^n+\eta_kz_{j,k}^n+\gamma_kr_{j,k}+\lambda_kg_{j,k}+\delta_km_{j,k},
\end{aligned}
\end{equation}
where $z_{j,k}^n$ is the destination indicator. If the $j$th charging station  is near the EV owner's travel destination (within a threshold distance $d_{th}$), $z_{j,k}^n=1$, otherwise, $z_{j,k}^n=0$. The term $d_{j,k}^n$ is the deviating distance due to EV charging. We use Dijkstra's shortest path algorithm \cite{Algorithm} to calculate the travel route for each EV owner from his/her origin to destination. If an EV owner needs to go to charging station $j$ halfway, we define the deviating distance $d_{j,k}^n$ as the route length of this new route minus the route length of the original route. Additionally, each candidate charging station has a vector of characteristics $[r_{j,k}, g_{j,k}, m_{j,k}]^{\mathrm{T}}$, which characterizes the attractiveness of this charging station in terms of those amenities. For instance, if there exists a restaurant near location $j$, we may set $r_{j,k} = 1$, otherwise we may set $r_{j,k} = 0$. Similarly, $g_{j,k}$ and $m_{j,k}$ are the indicators for shopping center and supermarket, respectively. The corresponding weighting coefficients are $\mu_k,\eta_k,\gamma_k,\lambda_k,\delta_k$.

The EV owner's choice is not deterministic due to the random unobservable utility. However, we can derive the average probability that s/he will choose a certain charging station by taking the expectation over the unobservable utilities defined in Equation (\ref{cdf}). The probability that the $n$-th EV owner will choose the $j$-th charging station of service provider $k$ is \cite{cite11}

\begin{equation}
\begin{aligned}
\Phi_{j,k}^n&=\mathbf{Prob}\left(\overline{U}_{j,k}^n+\epsilon_{j,k}^n > \overline{U}_{i,l}^n + \epsilon_{i,l}^n, \forall i \neq j, \textrm{or } l \neq k \right)\\
&=\int_{-\infty}^{+\infty}F_{j,k}(\overline{U}_{j,k}^n-\overline{U}_{1,1}^n+\epsilon_{j,k}^n,\overline{U}_{j,k}^n-\overline{U}_{2,1}^n+\epsilon_{j,k}^n,\\
&\cdots,\epsilon_{j,k}^n,\cdots,\overline{U}_{j,k}^n-\overline{U}_{L,3}^n+\epsilon_{j,k}^n)d\epsilon_{j,k}^n,
\end{aligned}
\end{equation}
where $F_{j,k}$ denotes the derivative of $F$ with respect to $\epsilon_{j,k}^n$, i.e. $F_{j,k}=\partial F/\partial \epsilon_{j,k}^n$. Evaluating this integral with the above assumptions, we obtain

\begin{equation}
\Phi_{j,k}^n=\frac{e^{\overline{U}_{j,k}^n/\sigma_k}\left(\sum_{l=1}^Le^{\overline{U}_{l,k}^n/\sigma_k}\right)^{\sigma_k-1}}
{\sum_{t=1}^3\left(\sum_{l=1}^Le^{\overline{U}_{l,t}^n/\sigma_t}\right)^{\sigma_t}}.
\end{equation}

\subsection{Charging Demand Estimation}
Once the EV owners' choice probability is computed, we can predict the charging demand at any charging station.  Let $q_n\;(n=1,2,\cdots,N)$ denote the total electricity (measured in kWh) that the $n$-th EV owner purchases from the charging station. Further, let $q_n$ be a random variable uniformly distributed in the range $[Q_a, Q_b]$, where $Q_a$ and $Q_b$ are, respectively, the lower and upper limit of charging demand for all EVs. The total predicted charging demand of charging station $j$ of service provider $k$ is modeled as
\begin{equation}
\psi_{j,k}=\sum_{n=1}^Nq_n\Phi_{j,k}^n.
\end{equation}

All coefficients in the nested logit model can be estimated and calibrated from preference survey data. The nested logit model enables us to compute the probability that an EV owner will go to a certain charging station, even though, an EV owner's decision may not always comply with the calculated probabilities. An individual EV owner may go to a fixed charging station at his/her discretion. However, employing the nested logit model provides a statistically meaningful prediction for the charging demand based on ensemble averages.

\section{The Impact of EV Integration on Power Grid}
The main function of the power grid is to deliver electricity to users reliably and economically. However, large-scale EV integration can potentially disrupt the normal operation of power grid in terms of system stability, severe power loss, frequency variation, voltage imbalance, etc. Generally, the variations in voltage and frequency of electricity are considered as the major factors to characterize the power quality.

Assume that the power system has $M$ generators and $D$ buses (substations), and the power flow study approach \cite{powerflow} is applied to solve the voltage, real power, and reactive power in the power system. Consider the node power equations, which can be written as real and reactive power for each bus.

\begin{equation}\label{networkflow1}
0=-P_i+\sum_{k=1}^N|v_i||v_k|(G_{ik}\cos\phi_{ik}+B_{ik}\sin\phi_{ik}),
\end{equation}
\begin{equation}\label{networkflow2}
0=-Q_i+\sum_{k=1}^N|v_i||v_k|(G_{ik}\sin\phi_{ik}-B_{ik}\cos\phi_{ik}),
\end{equation}
where $P_i$ and $Q_i$ are, respectively, the injected real power and reactive power at bus $i$. The variable $G_{ik}$ is the real part of the element in the bus admittance matrix corresponding to the $i$-th row and $k$-th column, and $B_{ik}$ is the imaginary part of the element. In Equations (\ref{networkflow1}) and (\ref{networkflow2}), $\phi_{ik}$ is the voltage angle between the $i$-th bus and the $k$-th bus, while $|v_i|$ and $|v_k|$ are the voltage magnitudes at bus $i$ and bus $k$, respectively. The stress on the voltage and frequency imposed by concentrated charging at EV charging station is expected to be significant. However, good models to calculate and penalize this stress are not yet considered.

In \cite{tong}-\cite{eric}, the authors have proposed different frameworks to coordinate EV charging to ensure stable and economical operation of power grid. In this work, we will consider how to alleviate the ``stress" added to the power grid by EV charging when determining the optimal charging station deployment.

Note that the voltage and system frequency are the two important factors of power quality in the power grid. The imbalance of the active power will lead to frequency drift, while the imbalance of the reactive power will cause the voltage variation \cite{powerstability}. Specifically, if the active power needed by the loads exceeds the generation, the extra active power is supplied by decreasing the generator's rotation speed, which results in a downward drift in frequency \cite{regulation}. On the other hand, the balance of the reactive power can influence the voltage stability. The excess of reactive power will lead to voltage increase, while the insufficiency of reactive power will lead to voltage decrease \cite{reactivepower}. Therefore, the demand and supply of both active power and reactive power should always be balanced.

Generally, the system operator schedules the power plants by estimating the load. If the load fluctuates significantly around the predefined value, the power plants are incentivized to ramp up and ramp down. However, this results in low efficiency and high cost from committing spinning reserves \cite{spinningreserve}. Therefore, the fluctuation of the active power and reactive power at the generators with and without EV charging can be used as a metric to evaluate the stress that the charging stations impose on the grid. In particular, we use the 2-norm deviation of generating power (real power and reactive power) of all generators in the power system to calculate the impacts of EV charging.

\begin{equation}
B=\norm{\mathbf{P}_{\textrm{g}}^{\textrm{base}}-\mathbf{P}_{\textrm{g}}^{\textrm{EV}}}^2_2+
\norm{\mathbf{Q}_{\textrm{g}}^{\textrm{base}}-\mathbf{Q}_{\textrm{g}}^{\textrm{EV}}}^2_2,
\end{equation}
where $\mathbf{P}_{\textrm{g}}^{\textrm{base}}=[P_1^{\textrm{base}},P_2^{\textrm{base}},\cdots,P_M^{\textrm{base}}]$ is a vector representing the active power generated by the $M$ generators under the base power load scenario (without EV charging), and $\mathbf{P}_{\textrm{g}}^{\textrm{EV}}=[P_1^{\textrm{EV}},P_2^{\textrm{EV}},\cdots,P_M^{\textrm{EV}}]$ is the vector of active power with EV charging (i.e., base power load superposed by EV charging load). Similarly, $\mathbf{Q}_{\textrm{g}}^{\textrm{base}}=[Q_1^{\textrm{base}},Q_2^{\textrm{base}},\cdots,Q_M^{\textrm{base}}]$ is the vector of reactive power of the base load, and $\mathbf{Q}_{\textrm{g}}^{\textrm{EV}}=[Q_1^{\textrm{EV}},Q_2^{\textrm{EV}},\cdots,Q_M^{\textrm{EV}}]$ is the vector of reactive power with EV charging. For a specific power system, $\mathbf{P}_{\textrm{g}}^{\textrm{base}},\mathbf{P}_{\textrm{g}}^{\textrm{EV}},\mathbf{Q}_{\textrm{g}}^{\textrm{base}}$, and $\mathbf{Q}_{\textrm{g}}^{\textrm{EV}}$ can be calculated through solving the global power flow equations.

\section{Spatial Competition and Optimal Placement in A Bayesian Game}
It is pivotal for firms to choose the right location and product to compete with rivals in the same industry. Business locations will affect business competition, and conversely intensive competition will affect how firms choose the appropriate locations. One question arises naturally is whether or not firms from the same industry like to cluster their stores. There are some classical literature on spatial competition, e.g. Hotelling's location model \cite{hotelling} and Salop's circle model \cite{outside2}. Firms have incentives for both clustering and separation. On one hand, firms prefer clustering so that they can learn from each other on how to improve manufacturing and research productivity \cite{glaeser}-\cite{shaver}, and learn demand from each other to reduce the cost of searching for the optimal location. Firms also cluster for the labor pool and supplies. In addition, firms can benefit from the spinoffs that are located near parent firms. On the other hand, the fear of intensive price competition due to clustering may motivate the firms to separate locations from each other.

The EV charging service providers face the similar dilemma. Therefore, we need to investigate how the service providers will interact with each other in choosing their charging station locations and setting the retail charging prices. In practice, the exact placement costs and utility functions of the competing service providers may not be known to the service provider \emph{a priori}. We thus formulate the problem as a Bayesian game \cite{bayesian} among the service providers at each planning stage.

For notational simplicity, we omit the stage superscript in the following definitions since the Bayesian game has the same mechanism at each stage. The main components of a Bayesian game include the set of players $\mathcal{I}$, the strategy space $S_k$, the type space $\Theta_k$, the payoff function $u_k$,  and the joint probability of the types $f(\Theta_1,\Theta_2,\Theta_3)$. $S_k=[s_{1,k},s_{2,k},\cdots, s_{L,k}]^{\textrm{T}}$ accounts for all possible placement policies for player $k$. $S_{-k}$ is the placement policies of the competing players. Denote $f(S_{-k})$ as probability mass function (PMF) of the placement policies of the other players. The type space $\Theta_k=[\theta_{1,k},\theta_{2,k},\cdots,\theta_{L,k}]^{\textrm{T}}$ corresponds to the placement costs of all charging station candidates of service provider $k$. In this paper, we assume that a service provider knows its own type, but not the types of the other two competing players.

Let $\theta_{j,k}$ denote the placement cost for $j$-th charging station candidate of service provider $k$, which includes the charging equipment cost, installation fee, construction cost, land rental cost, etc. $\theta_{j,k}$ has i.i.d. uniform distribution.

Before proceeding to analyze the Bayesian game, we need the following assumptions.

$\mathbf{Assumption\;1}$: $f(S_{-k})$ is binomially distributed with parameter 0.5, \emph{i.e.} $S_{-k}\backsim \textrm{Binomial}(2L, 0.5)$.

 $\mathbf{Remark\;1}$: The distribution of $S_{-k}$ reflects player $k$'s conjecture on how other players will act during the game. Each player can form their conjectures about other players according to their beliefs about the competitors. For instance, a player may be risk neutral, risk aversion or risk seeking. For simplicity, we assume that $S_{-k}$ has a binomial distribution with parameter 0.5. However, the theoretical analysis can be applied to any other distribution of $S_{-k}$.

$\mathbf{Assumption\;2}$: All service providers in the market are Bertrand competitors.

$\mathbf{Remark\;2}$: Bertrand competitors are players that do not cooperate with each other. Their goal is to maximize their own utility. They will not form any type of ``coalition" to manipulate the market.

For each player, the Bayesian Nash Equilibirum (BNE) of placement policy can be derived from Equation (\ref{obj}). To solve Equation (\ref{obj}), we need to know the retail charging prices of all the service providers. In a Bertrand competition, the retail prices for every combination of the charging station placement policies are determined by the first order of conditions (FOC):
\begin{equation}\label{eq1}
\frac{\partial\Pi_1}{\partial p_1}=\sum_{n=1}^N\sum_{j=1}^{L}q_ns_{j,1}\left[\Phi_{j,1}^n+(p_1-c_{j,1})\frac{\partial\Phi_{j,1}^n}{\partial p_1}\right]=0
\end{equation}
\begin{equation}\label{eq2}
\frac{\partial\Pi_2}{\partial p_2}=\sum_{n=1}^N\sum_{j=1}^{L}q_ns_{j,2}\left[\Phi_{j,2}^n+(p_2-c_{j,2})\frac{\partial\Phi_{j,2}^n}{\partial p_2}\right]=0
\end{equation}
\begin{equation}\label{eq3}
\frac{\partial\Pi_3}{\partial p_3}=\sum_{n=1}^N\sum_{j=1}^{L}q_ns_{j,3}\left[\Phi_{j,3}^n+(p_3-c_{j,3})\frac{\partial\Phi_{j,3}^n}{\partial p_3}\right]=0
\end{equation}
where $c_{j,1},c_{j,2},$ and $c_{j,3}$ represent the LMP at each charging station candidate.

$\mathbf{Remark\;3}$: For simulation simplicity, we assume the charging stations affiliated to the same service provider have the same retail charging prices ($p_1, p_2, \textrm{ and } p_3$). However, our analysis can be easily generalized to the case where each charging station sets its own retail price. Note that those retail prices obtained from Equations (\ref{eq1})-(\ref{eq3}) may not be the real-time prices used in practice. They are only the equilibrium prices in this market under the assumption of Bertrand competition. They can be interpreted as the averaged charging prices of the service providers over a long period of time. In practice, the service providers take turns to set the retail price in response to the prices of the competitors. Additionally, if some of the other factors change (i.e. consumer's preference, crude oil price soaring, etc.), the existing equilibrium breaks and a new equilibrium must be computed using the same procedure.

$\mathbf{Theorem\;1}$ [Strategy Decision Condition]: Service provider $k$ will choose placement policy $l(l=1,2,3,\cdots,2^L)$ if the type space $\Theta_k$ falls into the hypervolume specified by
\begin{equation}
\begin{aligned}
&\mathcal{H}(l)=\\
&\{\Theta_k\in \mathbb{R}_+^L:\Theta_k^{\textrm{T}}\left(S_{k,j}-S_{k,l}\right)-(\mathbb{E}R_{k,j}-\mathbb{E}R_{k,l})\\
&\;\;\;\;\;\;\;\;\;\;\;\;\;\;\;\;+w(B_{k,j}-B_{k,l})>0;\forall j \neq l\},
\end{aligned}
\end{equation}
where $S_{k,j}$ and $S_{k,l}$ denote the placement strategy $j$ and $l$, respectively. $\mathbb{E}R_{k,j}$ and $\mathbb{E}R_{k,l}$ denote the expected total revenue with deployment strategy $j$ and $l$, respectively.

\begin{proof}
Each service provider has $L$ location candidates, so there are $2^L$ different placement policies. The type space can be seen as an $L$-dimensional space, and $\Theta_k=[\theta_{1,k},\theta_{2,k},\cdots,\theta_{L,k}]^{\textrm{T}}$ represents a point in this space.

By Equation (\ref{obj}), strategy $l$ is optimal if
\begin{equation}
\begin{aligned}
&\mathbb{E}[R_{k,l}]-\Theta_{k}^{\textrm{T}}S_{k,l}-wB_{k,l}>\mathbb{E}[R_{k,j}]-\\
&\;\;\;\;\;\;\;\;\;\;\Theta_{k}^{\textrm{T}}S_{k,j}-wB_{k,j};(j=1,2,\cdots,2^L, j \neq l).
\end{aligned}
\end{equation}
Rearranging the terms, we get
\begin{equation}
\begin{aligned}
&\Theta_k^{\textrm{T}}\left(S_{k,j}-S_{k,l}\right)-(\mathbb{E}R_{k,j}-\mathbb{E}R_{k,l})+\\
&\;\;\;\;\;\;\;w(B_{k,j}-B_{k,l})>0;(j=1,2,\cdots,2^L, j \neq l),
\end{aligned}
\end{equation}
where each inequality represents a hyperplane and the intersection of all the inequalities defines a hypervolume in the type space.
\end{proof}

\section{Simulation Platform and Case Study}
\begin{figure}[htbp]
\centerline{\includegraphics[width=2.8in]{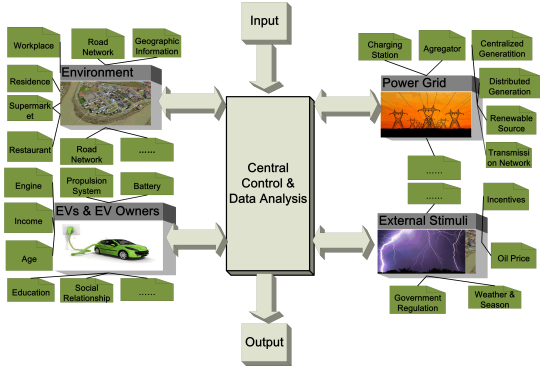}}
\center
\caption{The Architecture of The EV Virtual City 1.0}
\label{architecture}
\end{figure}

We have developed a general-purpose simulation software---The EV Virtual City 1.0 using Repast \cite{Repast}. Our simulation software is designed to construct a virtual digital city by integrating a variety of data and information, such as geographic information, demographic information, spatial infrastructure data, urban road network graph, electric power network graph, travel pattern, diurnal variation in traffic flow, seasonal fluctuation of driving activities, social interaction, etc. The platform is flexible in that one can include or exclude many modules to satisfy different simulation needs. See Fig. \ref{architecture} for the architecture of the simulation software.

\begin{figure}[htbp]
\centerline{\includegraphics[width=2.8in]{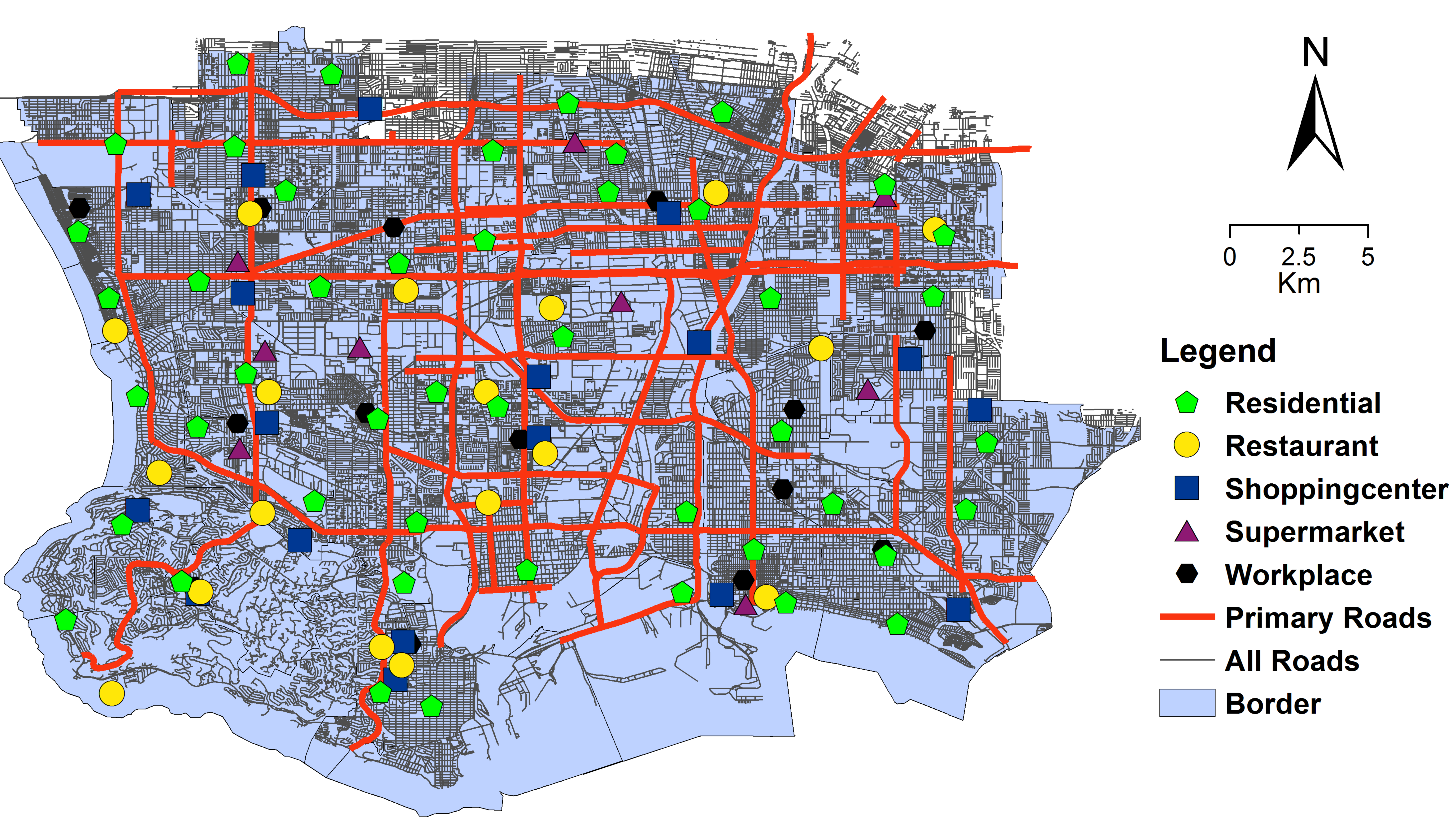}}
\center
\caption{Roads and Buildings of San Pedro District}
\label{road}
\end{figure}

In this paper, we conduct a case study using the data of San Pedro District of Los Angeles. We import the shapefiles of zip code tabulation area (ZCTA) and road network data from the U.S. Census Bureau into our simulation software. In addition, we calculate the centroids of locations of residence, restaurants, supermarkets, shopping centers and workplaces using Google Maps, see Fig. \ref{road}.

From the California Energy Commission website, we obtained the maps of transmission line and substations of San Pedro District. This area has 107 substations in total. Thus we use the IEEE 118-bus power system test case in our simulation. For each charging station placement policy, we used MATPOWER \cite{matpower} to calculate the LMP of each bus and the output power of each generator with and without EV charging.

\begin{figure}[htbp]
\centerline{\includegraphics[width=2.8in]{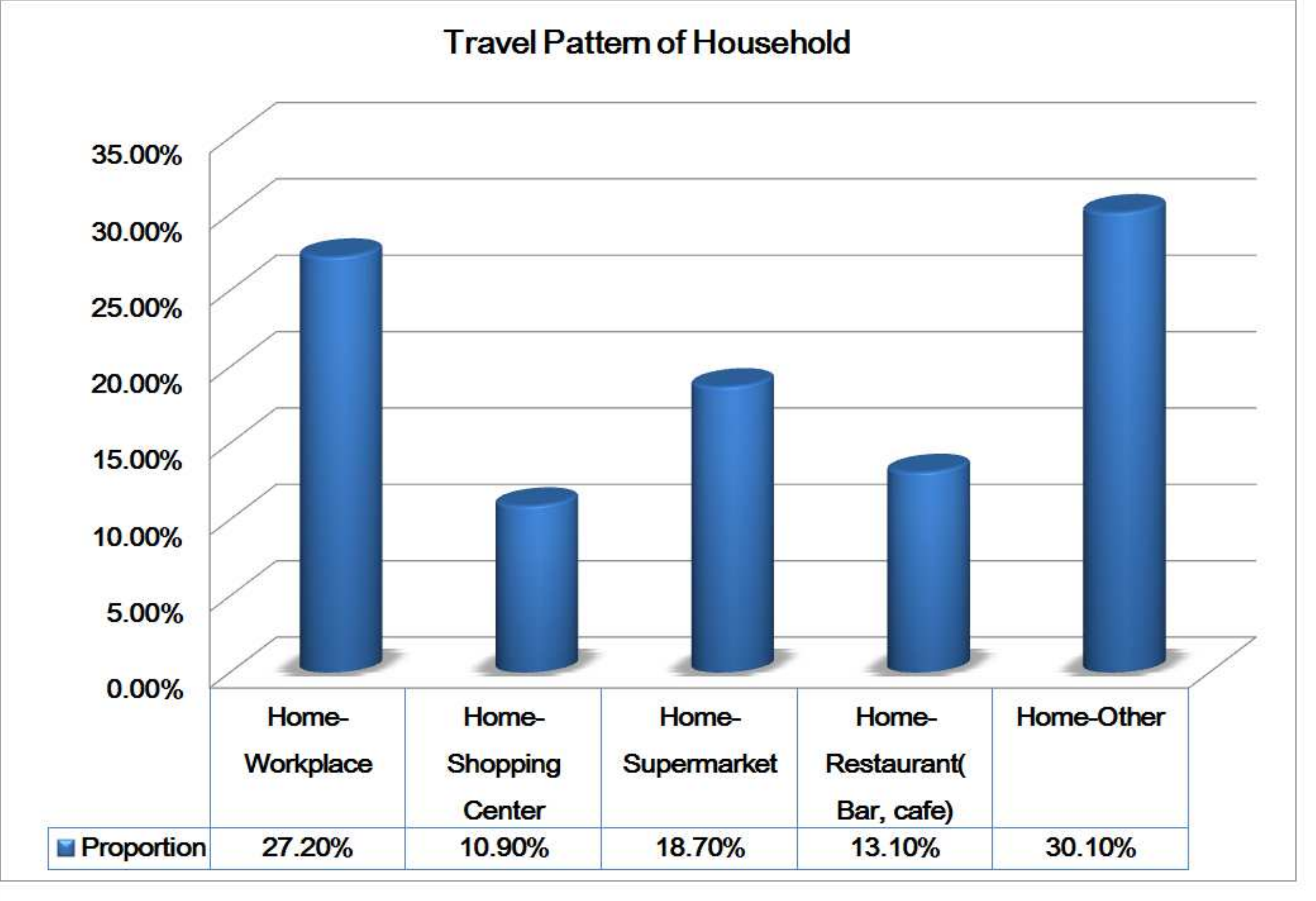}}
\center
\caption{The Statistics of Travel Patterns}
\label{statistic}
\end{figure}

The service providers must also satisfy the QoS constraints when planning the charging stations at each new stage. In the simulations, we consider four stages with 5000 EVs, 10000 EVs, 15000 EVs, and 20000 EVs. Since travel pattern also plays a significant role in analysing the charging behavior of EV owners, it is necessary to have a thorough study on the statistics of travel pattern. From the 2009 National Household Travel Survey (2009 NHTS) \cite{cite14}, we obtained the travel pattern statistics. See Fig. \ref{statistic}.

\begin{algorithm}{The Simulation Algorithm}\label{Algorithm1}
\begin{itemize}
    \item {\em Initialization:} Initialize road network, spatial infrastructure data, travel pattern statistics.

    \item {\em EV Movement:} For $1\leq j \leq N$ (total number of EVs), randomly assign a destination $Des(j)$ to EV $j$; calculate a route $Route(j)$ from $Home(j)$ to $Des(j)$ using Dijkstra's shortest path algorithm.

    \item {\em Bayesian Game Solution:}

        \begin{enumerate}
        \item {\em Prices and Revenue Calculation:} For $1 \leq i \leq M$ (total number of deployment strategies), calculate retail charging prices $p_1,p_2,\textrm{and } p_3$. Calculate $R_{1,i}, R_{2,i},\textrm{and},R_{3,i}$.
        \item{\em Optimal Deployment Strategy Search:} Find the strategy for each service provider $k(k=1,2,3)$
        $$S_k^*=l^*=\argmax_{l\in\{1,2,\cdots,M\}}\left\{R_{k,l}-\Theta_k^{\textrm{T}}S_{k,l}\right\}$$

    \end{enumerate}
    \item{\em Report:} $S_1^*,S_2^*,S_3^*,p_1^*,p_2^*,p_3^*$
\end{itemize}
\end{algorithm}

\begin{figure}[htbp]
\centerline{\includegraphics[width=2.8in]{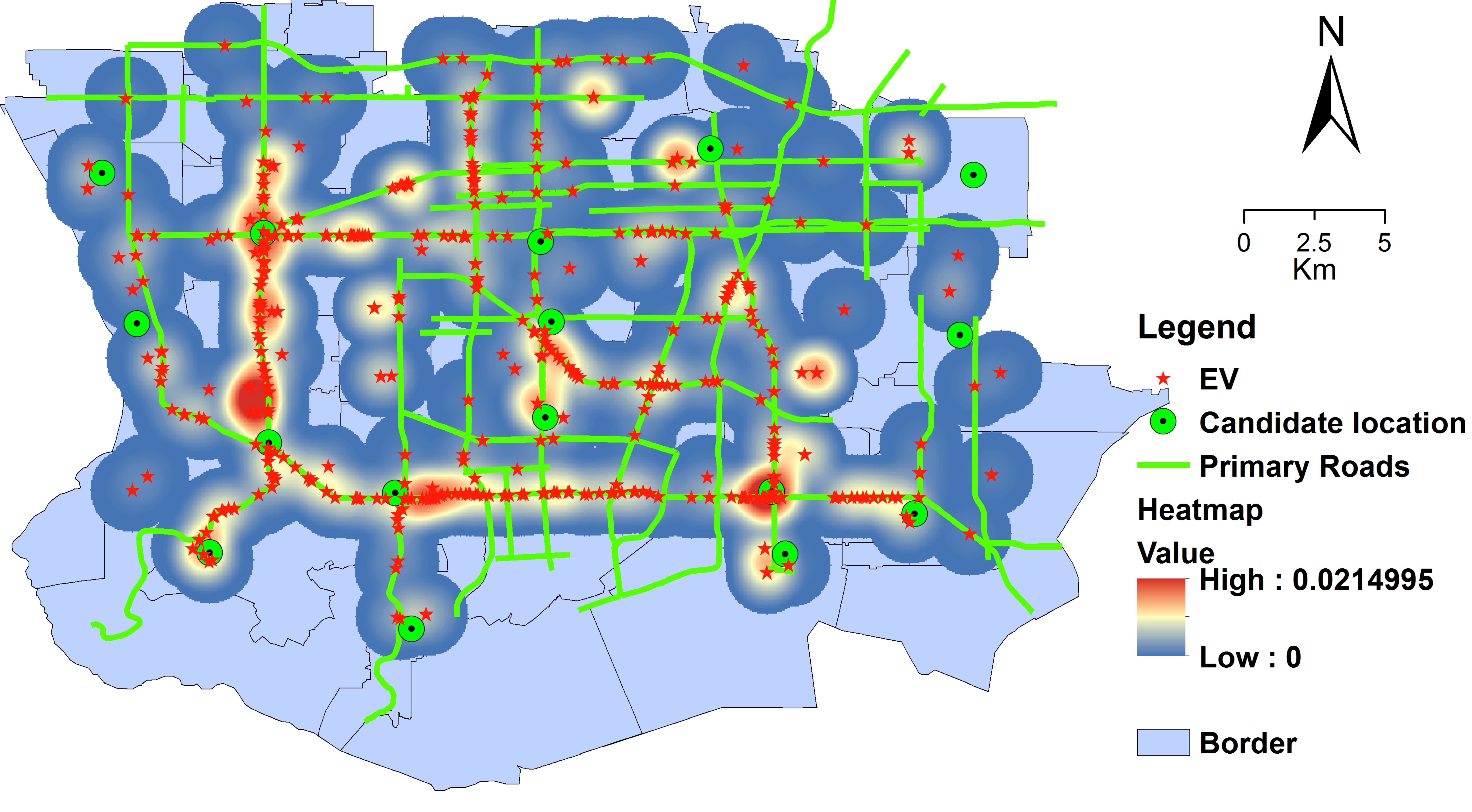}}
\center
\caption{A Snapshot of EVs Movement}
\label{movement}
\end{figure}

\begin{figure*}[htbp]
  \begin{minipage}[b]{0.6\linewidth}
    \includegraphics[width=2.7in]{newStage1Result-eps-converted-to.pdf}
    \caption{Stage 1 Charging Station Placement}
    \label{stage1}
  \end{minipage}
  \begin{minipage}[b]{0.6\linewidth}
    \includegraphics[width=2.7in]{newStage2Result-eps-converted-to.pdf}
    \caption{Stage 2 Charging Station Placement}
    \label{stage2}
  \end{minipage}
  \begin{minipage}[b]{0.6\linewidth}
    \includegraphics[width=2.7in]{newStage3Result-eps-converted-to.pdf}
    \caption{Stage 3 Charging Station Placement}
    \label{stage3}
  \end{minipage}
  \hfill
  \begin{minipage}[b]{0.6\linewidth}
    \includegraphics[width=2.7in]{newStage4Result-eps-converted-to.pdf}
    \caption{Stage 4 Charging Station Placement}
    \label{stage4}
  \end{minipage}
      \begin{minipage}[b]{0.6\linewidth}
    \includegraphics[width=2.7in]{finalResult-eps-converted-to.pdf}
    \caption{Superimposing all stages}
    \label{overview}
  \end{minipage}
  \begin{minipage}[b]{0.6\linewidth}
    \includegraphics[width=2.5in]{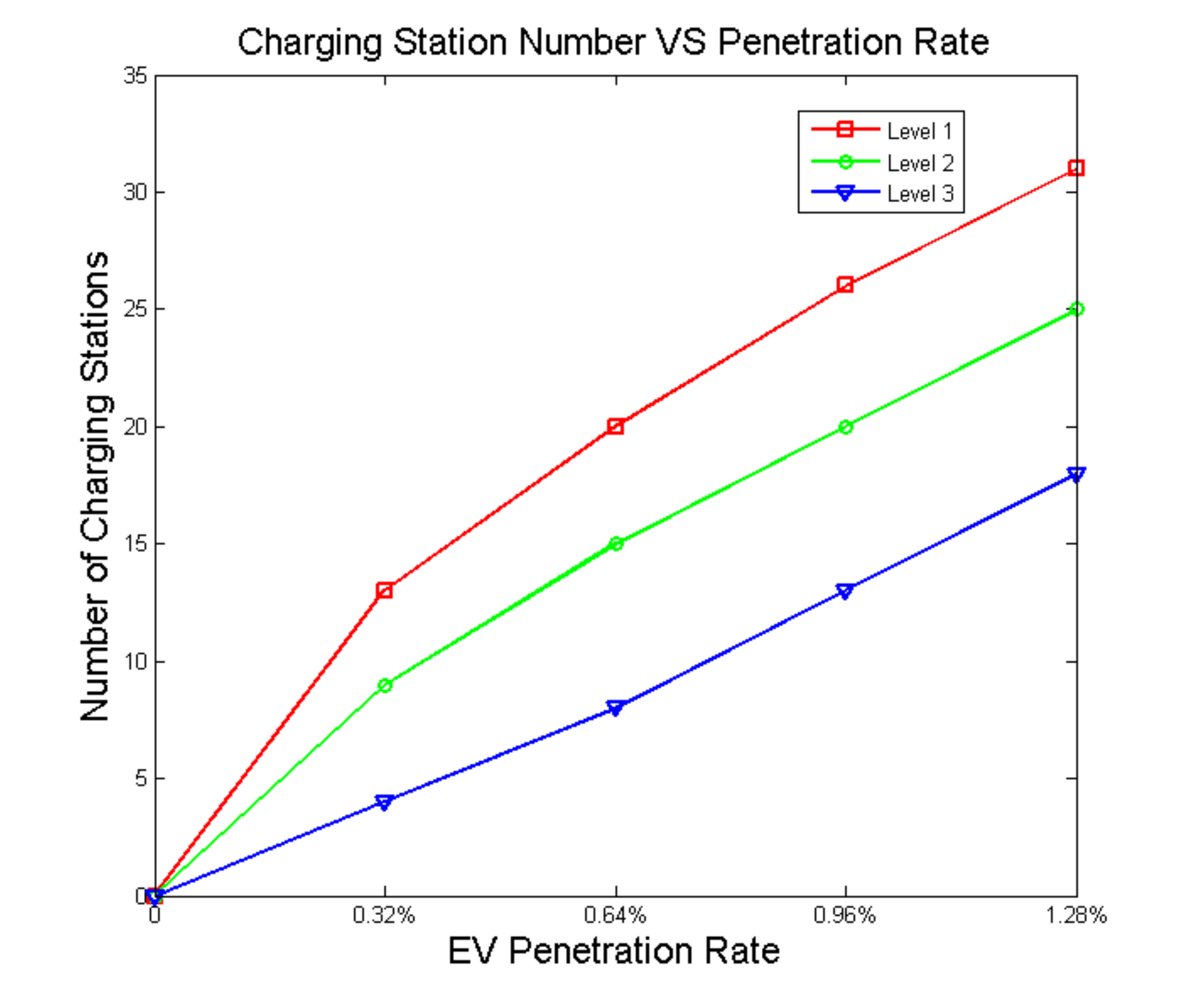}
    \caption{Charging Station verses EV Penetration Rate}
    \label{line}
  \end{minipage}
\end{figure*}

A snapshot of the moving EVs is shown in Fig. \ref{movement}. Each red star represents an EV owner. The traffic flow heatmap of EV owners is also plotted in this figure. The simulation results are summarized in Table \ref{ta9}. Figs. \ref{stage1} to \ref{stage4} correspond to the charging station placement for stages 1 to 4, respectively. Fig. \ref{overview} is an overview of charging station placement by superposing Figs. \ref{stage1} to \ref{stage4}. Fig. \ref{line} shows how the number of charging station increases as the EV penetration rate increases.

\begin{table*}[htbp]
\center
\caption{Charging Station Placement Strategy}\label{ta9}
\begin{tabular}{cccclc}
\toprule
Stage & Level & Delay prob. & Coverage & Newly Built Stations&Total \# of Stations\\
\midrule
Stage 1&1&0.29 & 1.7 & 1, 2, 3, 4, 5, 6, 7, 8, 9, 11, 12, 13, 14 & 13\\
(Penetration Rate 0.32\%, & 2& 0.17 & 1.6 &  2, 3, 4, 7, 9, 10, 12, 14, 15 & 9 \\
5000 EVs) & 3&0.04&0.8 & 2, 4, 7, 12 & 4\\
  \bottomrule
Stage 2 & 1&0.28 & 2.5 & 17, 20, 22, 23, 24, 25 & 20\\
(Penetration Rate 0.64\%, & 2& 0.17 & 2.3 &  18, 20, 22, 23, 24, 25 &15 \\
10000 EVs)&3&0.10&1.15 & 18, 22, 23, 25 & 8\\
\bottomrule
Stage 3 & 1&0.29 & 3.6 & 26, 27, 29, 30, 31, 32 & 26\\
(Penetration Rate 0.96\%,  & 2& 0.18 & 3.1 & 26, 27, 28, 31, 32 & 20\\
15000 EVs)&3&0.08&1.9 & 26, 27, 28, 29, 31 & 13\\
\bottomrule
Stage 4 & 1&0.22 & 4.6 & 33, 34, 35, 36, 37 & 31\\
(Penetration Rate 1.28\%,  & 2& 0.16 & 3.9 & 33, 37, 38, 39, 44 & 25\\
20000 EVs)&3&0.10&2.9 & 36, 38, 40, 41, 44 & 18\\
\bottomrule
\end{tabular}
\end{table*}

From the simulation results, we can make the following observations:
\begin{itemize}
\item The optimal charging station deployment is consistent with the EV traffic flow heatmap. This suggests that our model can adequately capture the mobility of EVs and provide EV owners with convenient charging services.

\item  As for the number of charging stations, Level 1 charging station is predominant over Level 2 and Level 3. Level 3 has the least number of charging stations. Notice that it takes a much longer time to finish charging for Level 1, so Level 1 service provider must place more charging stations to meet the average delay probability constraint. The difference in quantity also reveals that the service providers have different marketing strategies. Service provider 1 tries to place the charging stations evenly across the entire area, while service provider 3 is more likely to place the charging stations at some ``hot" locations.

\item The number of charging stations grows almost linearly with the number of EVs except for the initial stage. At the initial stage, Level 1 and Level 2 service providers tend to place more charging stations than the next stages. This is because service providers must place more charging stations to meet the average service coverage constraints. As the number of charging stations increases, however, the service coverage constraint is less of a concern for the service providers.

\item Service providers prefer clustering instead of spatial separation. The three service providers have segmented the EV charging market by providing three different products (different charging level services) in terms of voltage, current, charging speed and charging price. Due to product differentiation, they significantly soften the price competition so that they do not need to spatially separate from each other to further relax competition. This observation supports the opinions in \cite{Competition}-\cite{chargingstation} that firms do not have to maximize differentiation in every characteristic of the product. In general, differentiation in one dominant characteristic is sufficient to soften price competition.
\end{itemize}

\section{Conclusion}
In this paper, we have proposed a solution to the placement of EV charging stations which balances the benefits of EV owner, charging station owner, and power grid operator. We formulate the competition among charging stations as a Bayesian game. Solving the game renders the optimal placement policies for the service providers. In addition, we develop a simulation software---The EV Virtual City 1.0 on Repast, and conduct a case study of San Pedro District of Los Angeles. The simulation illustrates that charging station placement is highly consistent with the traffic flow of EVs, and the service providers prefer clustering to separating the charging stations.

\begin{IEEEbiography}[{\includegraphics[width=1.1in,height=1.2in,clip,keepaspectratio]{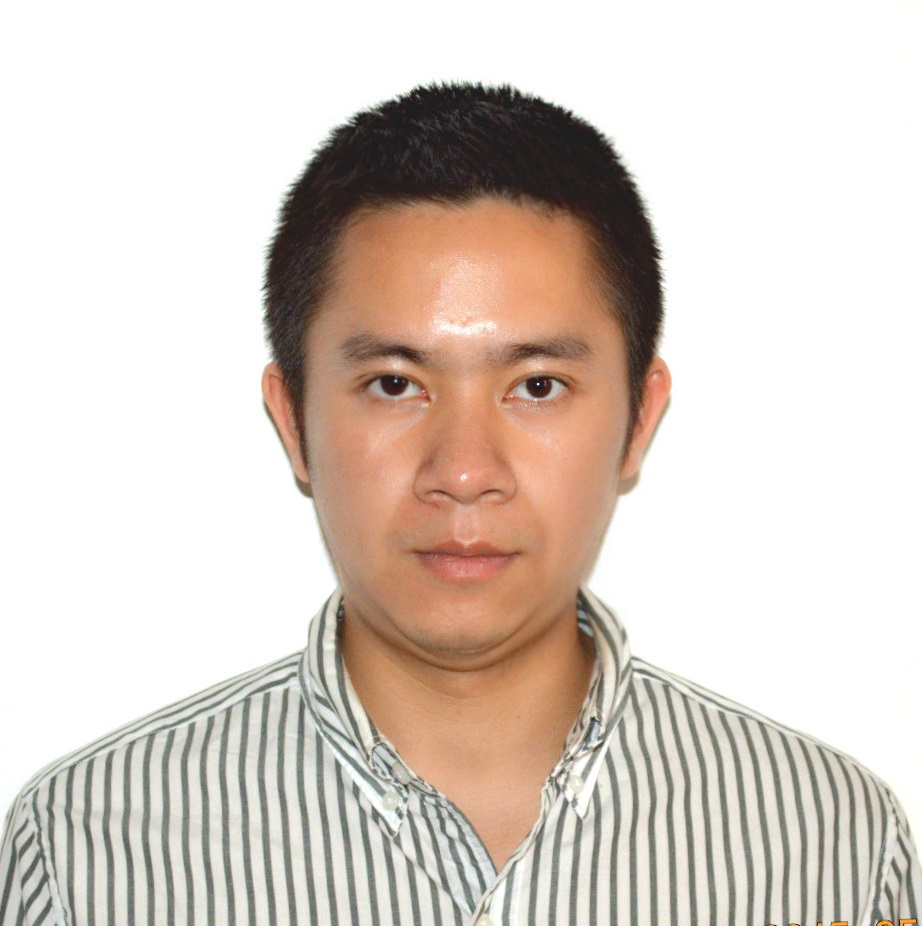}}]{Chao Luo}
received the B.Eng degree with distinction in Communication Engineering from Harbin Institute of Technology (HIT), China, in 2012. Currently, he is pursuing his Ph.D. in Electrical Engineering in the University of Notre Dame, USA. Chao's research interests include electric vehicle (EV) integration into power grid, network protocol design of smart grid, and electricity market.
\end{IEEEbiography}

\begin{IEEEbiography}[{\includegraphics[width=1in,height=1.25in,clip,keepaspectratio]{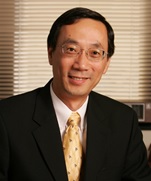}}]{Yih-Fang Huang}
is Professor of Department of Electrical Engineering and Senior Associate Dean of College of Engineering at University of Notre Dame.  Dr. Huang received his BSEE degree from National Taiwan University, MSEE degree from University of Notre Dame and Ph.D. from Princeton University. He served as chair of the Electrical Engineering Department at the University of Notre Dame from 1998 to 2006.  His research interests focus on theory and applications of statistical signal detection and estimation, and adaptive signal processing.

In Spring 1993, Dr. Huang received the Toshiba Fellowship and was Toshiba Visiting Professor at Waseda University, Tokyo, Japan.  From April to July 2007, he was a visiting professor at the Munich University of Technology, Germany.  In Fall, 2007, Dr. Huang was awarded the Fulbright-Nokia scholarship for lectures/research at Helsinki University of Technology in Finland (which is now Aalto University).

Dr. Huang received the Golden Jubilee Medal of the IEEE Circuits and Systems Society in 1999, served as Vice President in 1997-98 and was a Distinguished Lecturer for the same society in 2000-2001.  At the University of Notre Dame, he received Presidential Award in 2003, the Electrical Engineering department's Outstanding Teacher Award in 1994 and in 2011, the Rev. Edmund P. Joyce, CSC Award for Excellence in Undergraduate Teaching in 2011, and the College of Engineering's Teacher of the Year Award in 2013.  Dr. Huang is a Fellow of the IEEE.
\end{IEEEbiography}

\begin{IEEEbiography}[{\includegraphics[width=1in,height=1.25in,clip,keepaspectratio]{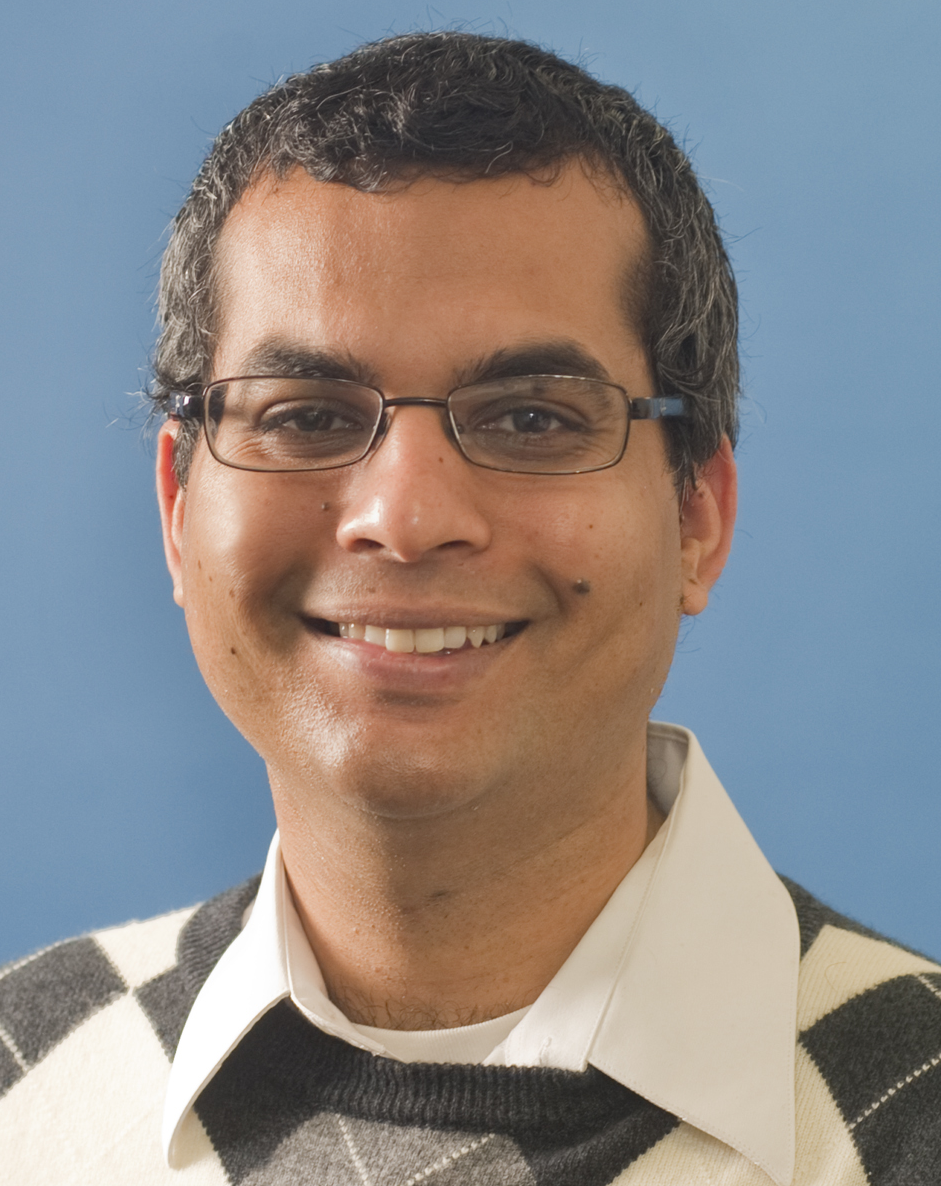}}]{Vijay Gupta}
is an Associate Professor in the Department of Electrical Engineering at the University of Notre Dame. He received his B. Tech degree from the Indian Institute of Technology, Delhi and the M.S. and Ph.D. degrees from the California Institute of Technology, all in Electrical Engineering. Prior to joining Notre Dame, he also served as a research associate in the Institute for Systems Research at the University of Maryland, College Park. He received the NSF CAREER award in 2009, and the Donald P. Eckman Award from the American Automatic Control Council in 2013. His research interests include cyber-physical systems, distributed estimation, detection and control, and, in general, the interaction of communication, computation and control.
\end{IEEEbiography}






\end{document}